\documentstyle[12pt]{article}
\newcommand{\beq}{\begin{equation}}
\newcommand{\eeq}{\end{equation}}
\newcommand{\bea}{\begin{eqnarray}}
\newcommand{\eea}{\end{eqnarray}}

\begin{document}
\setcounter{page}{0}
\topmargin 0pt
\oddsidemargin 5mm
\renewcommand{\thefootnote}{\fnsymbol{footnote}}
\newpage
\setcounter{page}{0}
\begin{titlepage}
\begin{flushright}
QMW 94-24
\end{flushright}
\begin{flushright}
hep-th/94mmxxx
\end{flushright}
\vspace{0.5cm}
\begin{center}
{\large {\bf Solutions of the Master Virasoro Equation as Conformal Points of
Perturbed WZNW Model}} \\
\vspace{1.8cm}
\vspace{0.5cm}
{\large Oleg A. Soloviev
\footnote{e-mail: soloviev@V1.PH.QMW.ac.uk}\footnote{Work supported by
PPARC and in part by a contract from the European Commission Human Capital and
Mobility Programme.}}\\
\vspace{0.5cm}
{\em Physics Department, Queen Mary and Westfield College, \\
Mile End Road, London E1 4NS, United Kingdom}\\
\vspace{0.5cm}
\renewcommand{\thefootnote}{\arabic{footnote}}
\setcounter{footnote}{0}
\begin{abstract}
{It is shown that the master equation of the affine-Virasoro
construction on the unitary affine algebra naturally emerges in the
fusion algebra of the nonunitary level $k$
WZNW model. Operators corresponding
to solutions of the master equation are suitable for performing
one-parametrical renormalizable
perturbation around the given conformal nonunitary WZNW model. In the large
$|k|$ limit, the infrared fixed point of the renormalization group beta
function is found. There are as many infrared conformal points as
$\frac{1}{2}N(N+1)$, where $N$ is the total number of solutions of
the master equation.}
\end{abstract}
\vspace{0.5cm}
\centerline{August 1994}
 \end{center}
\end{titlepage}
\newpage
%******************************************************************
\section{Introduction}

There is a large class of two dimensional CFT's which are described by the
affine-Virasoro construction \cite{Kiritsis},\cite{Morozov}. The latter is the
most general bilinear of the affine currents. A study of embeddings of the
affine-Virasoro construction into the affine
algebra has resulted in a master equation \cite{Kiritsis},\cite{Morozov},
particular solutions of which correspond to the affine-Sugawara
\cite{Bardakci-1},\cite{Halpern},\cite{Knizhnik}, coset
\cite{Bardakci-1},\cite{Halpern},\cite{Goddard}, and spin-orbit
\cite{Kiritsis},\cite{Bardakci-1},\cite{Mandelstam} constructions. Being
exactly solvable the aforementioned conformal theories admit proper Lagrangian
formulations. There has been growing interest, recently,
in the question of whether or not all other solutions (emeddings) of the
master equation (affine-Virasoro construction) can have natural Lagrangian
descriptions \cite{Yamron}-\cite{Bardakci-2}.

A classical action for generic affine-Virasoro construction has been obtained
in \cite{Yamron} by utilizing the Hamiltonian approach to the
Wess-Zumino-Novikov-Witten (WZNW) model \cite{Bowcock}. Whereas in
\cite{Soloviev-1} an affine-Virasoro action is built up in terms of the
conformal
non-Abelian Thirring model. As yet it is not clear whether there
is a link between these
two actions. We have shown explicitly the conformal symmetry of the non-Abelian
Thirring model at the isoscalar Dashen-Frishman conformal point \cite{Dashen}
which turns out to be one of the solutions of the master Virasoro equation.
However, it is very difficult to arrive at all other conformal points starting
from the nonconformal Thirring model. Such an attempt has been done in ref.
\cite{Bardakci-2}. Finally ref. \cite{Tseytlin} was an effort to get the master
equation from the vanishing of renormalization group beta functions of a
certain nonlinear sigma model. Unfortunately, the last approach does not lead
beyond the classical solutions of the master equation. This result was somewhat
programmed by the use of the $1/k$-expansion method, where $k$ is level of the
underlying affine algebra. Indeed, in the large $k$
limit, the master equation collapses to the classical approximation
\cite{Obers} which cannot give more than classical solutions.

The aim of the present paper is to suggest another application of the master
equation in the quest for new conformal lagrangian field theories. We will
describe a situation in which the master equation emerges not as a condition of
the conformal
symmetry but as a condition of renormalizability of a certain quantum field
theory. Additional requirement of conformal invariance will result in an
equation for the {\it one}-parametrical coupling constant.

Our approach to a certain extent is a generalization of the
idea of perturbing the
nonunitary WZNW model by its kinetic term \cite{Soloviev-2}. The latter
appears to be an appropriate relevant unitary Virasoro primary operator
in the spectrum of the nonunitary model \cite{Soloviev-2}.
Certainly, the nounitary WZNW model is a highly nontrivial theory. Because it
has states with negative norm, its proper definition is quite complicated and
has not been yet done completely. Therefore, part of our discussion may seem to
be somewhat formal. However, we will argue that all operators and manipulations
on them which we will use in this paper are unitary and well defined. We
believe that the nonunitary WZNW model itself will be finally properly
understood.

The paper is organized as follows. In section 2 the master Virasoro equation
associated with the affine-Virasoro construction is derived from the fusion
algebra of the nonunitary WZNW model. In section 3 the operators, whose fusion
algebra gives rise to the master equation, are utilized to perturb the
nonunitary WZNW model. We show that the perturbed theory has as many
infrared conformal points as $\frac{1}{2}N(N+1)$, where $N$ is the total number
of solutions to the master equations.

\section{Master Virasoro equation from nonunitary WZNW model}

Let us consider the level $k$ WZNW model whose action is given as follows
\cite{Witten}
\begin{equation}
S(k)=-{k\over4\pi}\{\int
\mbox{Tr}|g^{-1}\mbox{d}g|^2~+~{i\over3}\int\mbox{d}^{-1}\mbox{Tr}(g^{-1}
\mbox{d}g)^3\},\end{equation}
where $g$ is the matrix field taking its values on the Lie group $G$. At this
point we do not specify whether $k$ is positive or negative. The theory
possesses the affine symmetry $\hat G\times\hat G$ which entails an infinite
number of conserved currents \cite{Knizhnik},\cite{Witten}. The latter can be
derived from the basic currents $J$ and $\bar J$,
\begin{equation}
J=J^at^a=-{k\over2}g^{-1}\partial g,~~~~~~~~~\bar J=\bar
J^at^a=-{k\over2}\bar\partial gg^{-1},\end{equation}
satisfying the equations of motion
\begin{equation}
\bar\partial J=0,~~~~~~~~~~\partial\bar J=0.\end{equation}
In eqs. (2) $t^a$ are the generators of the Lie algebra ${\cal G}$ associated
with the Lie group $G$,
\begin{equation}
[t^a,t^b]=f^{abc}t^c,\end{equation}
with $f^{abc}$ the structure constants.

The spectrum of the WZNW model contains states which correspond to the primary
fields of the underlying affine symmetry \cite{Knizhnik}. By definition,
$\phi_i$ is an affine (chiral) primary field, if it has the following operator
product expansion (OPE) with the affine current $J$ \cite{Knizhnik}
\begin{equation}
J^a(w)\phi_i(z)={t^a_i\over w-z}\phi_i(z)~+~reg.,\end{equation}
where the matrices $t_i^a$ correspond to the (left) representation of
$\phi_i(z)$. In the WZNW model, any affine primary field is Virasoro primary
and its conformal dimension is given by \cite{Knizhnik}
\begin{equation}
\Delta_i={c_i\over c_V+k},\end{equation}
where $c_i=t^a_it^a_i$ and $c_V$ is defined according to
\begin{equation}
f^{acd}f^{bcd}=c_V~\delta^{ab}.\end{equation}

There are Virasoro primary states in the spectrum of the WZNW model which are
the descendants of the affine primary vectors. One of such states was
considered in \cite{Soloviev-1},\cite{Soloviev-2}. In what follows we are going
to describe a class of such fields.

Let us take the following composite field
\begin{equation}
O^{L,\bar L}=L_{ab}~\bar L_{\bar a\bar b}~:J^a\bar J^{\bar a}\phi^{b\bar b}:
\end{equation}
which is defined as a normal ordered product of the affine currents $J^a,~\bar
J^{\bar a}$ with the isospin (1,1) affine-Virasoro primary field $\phi^{b\bar
b}$ in the adjoint representation of $G\times G$. The product of the three
operators in eq. (8) can be properly defined according to
\begin{equation}
O^{L,\bar L}(z,\bar z)=L_{ab}\bar L_{\bar a\bar b}~\oint{dw\over2\pi
i}~\oint{d\bar w\over2\pi i}{J^a(w)\bar J^{\bar a}(\bar w)~\phi^{b\bar
b}(z,\bar z)\over|z-w|^2},\end{equation}
where the product in the numerator of the integrand is understood as an OPE.
It is easy to see
that the given product does not contain singular terms provided the matrices
$L_{ab}$ and $\bar L_{\bar a\bar b}$ are {\it symmetrical}.

{}From the definition it follows that the operator $O^{L,\bar L}$ is an affine
descendant of the affine-Virasoro primary field $\phi$. Indeed, $O^{L,\bar L}$
can be presented in the form
\begin{equation}
O^{L,\bar L}(0) =L_{ab}\bar L_{\bar a\bar b}~J^a_{-1}\bar J^{\bar
a}_{-1}\phi^{b\bar b}(0),\end{equation}
where
\begin{equation}
J^a_m=\oint{dw\over2\pi i}w^mJ^a(w),~~~~~~~\bar J^{\bar
a}_m=\oint{d\bar w\over2\pi i}\bar w^m\bar J^{\bar a}(\bar
w).\end{equation}
Being an affine descendant, the operator $O^{L,\bar L}$ continues to be a
Virasoro primary operator. Indeed, one can check that the state $O^{L,\bar
L}(0)|0\rangle$ is a highest weight vector of the Virasoro algebra, with
$|0\rangle$ the $SL(2,C)$ invariant vacuum. That is,
\begin{eqnarray}
L_0O^{L,\bar L}(0)|0\rangle&=&\Delta_O~O^{L,\bar L}(0)|0\rangle,\nonumber \\ &
&
\\
L_{m>0}O^{L,\bar L}(0)|0\rangle&=&0.\nonumber\end{eqnarray}
Here the generators $L_n$ are given by the contour integrals
\begin{equation}
L_n=\oint{dw\over2\pi i}~w^{n+1}T(w),\end{equation}
where $T(w)$ is holomorphic component of the Sugawara stress tensor of the
conformal WZNW model,
\begin{equation}
T(z)={:J^a(z)J^a(z):\over k+c_V}.\end{equation}
In eqs. (12), $\Delta_O$ is the conformal dimension of the operator $O^{L,\bar
L}$. We find
\begin{equation}
\Delta_O=\bar\Delta_O=1+{c_V\over k+c_V}.\end{equation}
Here $\bar\Delta_O$ is the conformal dimension of $O^{L,\bar L}$ associated
with
antiholomorphic conformal transformations. In what follows, we will discuss
the WZNW model with negative level. This theory has states with
negative norm. Remarkably the given operator
$O^{L,\bar L}$ corresponds to a unitary highest weight vector of the Virasoro
algebra when $|k|>c_V$ \cite{Soloviev-3}. Therefore, all correlation functions
of this operator make sense even when $k$ is negative.

Clearly, operators $O^{L,\bar L}$  with arbitrary symmetrical matrices
$L_{ab},~\bar L_{\bar a\bar b}$ are Virasoro primary vectors with the same
conformal dimensions. However, their fusion algebras may be different. We would
like to focus on a particular subclass of operators $O^{L,\bar L}$ which obey
the following fusion
\begin{equation}
O^{L,\bar L}\cdot O^{L,\bar L}=[O^{L,\bar L}]~+~[I]~+...\end{equation}
where the square brackets denote the contributions of $O^{L,\bar L}$ and
identity operator $I$ and the corresponding descendants of $O^{L,\bar L}$ and
$I$, whereas dots stand for all other admitted
operators with different conformal
dimensions. It is important to emphasize that the operator $O^{L,\bar L}$ on
the left and right hand sides of eq. (16) has one and the same pairs of
matrices
$L_{ab}$ and $\bar L_{\bar a\bar b}$. We will show that eq. (16) leads to
algebraic equations for the given matrices. These equations will be the main
subject of this section.

Due to the theorem of holomorphic factorization, we can forget for a while
about antiholomorphic part of the operator $O^{L,\bar L}$. First of all, we
have to compute the following OPE
\begin{equation}
\phi^a(w)~\phi^b(z)={\sum_I}(w-z)^{\Delta_I-2\Delta_\phi}~C_I^{ab}[\Phi^I(z)],
\end{equation}
where $[\Phi^I]$ are conformal classes of all Virasoro primaries $\Phi^I$
arising in the fusion of two $\phi$'s. For our purposes, it is sufficient to
calculate $C^{ab}_c~[\phi^c]$.

Let us set $z$ to zero in eq. (17). Then after acting on the $SL(2,C)$
vacuum $|0\rangle$, eq. (17) gives
\begin{equation}
\phi^a(w)|\phi^b\rangle=w^{-\Delta_\phi}C^{ab}_c~[|\phi^c\rangle]~+~...
\end{equation}
The structure constants $C^{ab}_c$ and the terms of $[|\phi^c\rangle]$
can be deduced
from the invariance of eq. (18) under the affine symmetry. Indeed, by acting
with $J_0^a$ and $J_1^a$ on both sides of eq. (18), we find
\begin{equation}
C^{ab}_c~[|\phi^c\rangle]=A\{f^{abc}|\phi^c\rangle~+~w~J^a_{-1}|\phi^b\rangle~
+~...\},\end{equation}
where $A$ is an overall constant whose value is not essential for our
consideration. It is instructive to verify that for $G=SU(2)$ and positive $k$
the same expression (19) can be derived by using the parafermion representation
\cite{Fateev} of the affine primary $\phi^a$ which corresponds to the field
$\Phi^J$ with $J=1$ in notations of ref. \cite{Fateev}. Besides, in the limit
$c_V\to\infty$, the field $\phi^a$ acquires dimension (1,0) of the affine
current. Note that
\begin{equation}
\Delta_\phi={c_V\over k+c_V}.\end{equation}
Apparently, eq. (19) is consistent with this limit.

All in all, we arrive at the following formula for the OPE of two $\phi$'s
\begin{equation}
\phi^a(w)~\phi^b(z)={[I]\over(w-z)^{2\Delta_\phi}}~+~A\left\{ {f^{abc}\over
(w-z)^{\Delta_\phi}}\phi^c(z)~+~{1\over(w-z)^{\Delta_\phi-1}}J^a_{-1}\phi^b(z)
{}~+~...\right\}~+~...\end{equation}
The point to be made is that in the
case $G=SL(n)$, the coefficient $A$ in eq. (21) can be fixed explicitly
using the free field representation method \cite{Dotsenko},\cite{Felder}.

Taking into account eq. (21), we obtain for the holomorphic part of the
operator
$O^{L,\bar L}$ the following fusion
\begin{eqnarray}
O(w)~O(z)&=&{[I]\over(w-z)^{2\Delta_O}}
{}~+~A{L_{ab}~L_{mn}\over(w-z)^{\Delta_O}}(\frac{k}{2}
\delta^{am}~J_{-1}^b\phi^n(z)~+~
f^{amk}f^{bnc}~J^k_{-1}\phi^c(z)\nonumber\\ & & \\
&+&f^{amk}f^{kbc}~J_{-1}^c\phi^n(z)~+~
f^{amk}f^{knc}~J_{-1}^b\phi^c(z))~+~...\nonumber\end{eqnarray}
Here dots stand for conformal operators with different conformal dimensions.

{}From formula (22) one can see that in general two operators $O^{L,\bar L}$
fuse into another operator $O^{\tilde L,\bar{\tilde L}}$.  Also, it
becomes clear that in order to have
$O^{L,\bar L}$ on the right hand side of eq. (22) with the same matrices
$L,~\bar L$, we have to impose the following condition
\begin{equation}
L_{ab}=-\frac{k}{2}
{}~L_{ac}L_{cb}~-~L_{cd}L_{ef}f^{cea}f^{dfb}~-~L_{cd}f^{cef}f^{dfa}
L_{be}~-~L_{cd}f^{cef}f^{dfb}L_{ae},\end{equation}
whith a similar equation for $\bar L_{\bar a\bar b}$.
It is rather amazing that the given equation is almost identical to the
celebrated master Virasoro equation \cite{Kiritsis},\cite{Morozov}. The only
difference is in the sign of the first term on the right hand side of (23). So,
if we
have dealt with the operator $O^{L,\bar L}$ associated with the unitary affine
algebra, then the equation we would get is the master equation of the
nonunitary affine-Virasoro construction. Whereas {\it
the master Virasoro equation for the unitary affine-Virasoro construction
originates from the fusion algebra of the operator $O^{L,\bar L}$
corresponding to the nonunitary
affine algebra.} This will be a crucial observation for the further discussion.
In particular, in the case of the unitary affine algebra, the existence of
Virasoro primary operators with
the fusion given by eq. (16) depends on the
existence of solutions to the nonunitary master Virasoro equation. While in the
case of the nonunitary affine algebra, there are definitely Virasoro primary
operators $O^{L,\bar L}$ in the spectrum of the nonunitary WZNW model,
obeying the fusion in eq. (16).  In other words, we
have shown that the unitary master equation resides in the fusion algebra of
the nonunitary WZNW model.

\section{$O^{L,\bar L}$-perturbation}

It is obvious that in the
limit $k\to-\infty$, $O^{L,\bar L}$  becomes a relevant
quasimarginal operator. Indeed, in this limit, the conformal dimension
$\Delta_O$ is just slightly less than one. This operator
will satisfy the fusion algebra given by eq. (16)
provided the
algebraic equation (23) is fulfilled. In order to use this operator as
a perturbation around the nonunitary WZNW model, we should be aware of the fact
that there
are no other relevant operators on the right hand side of eq. (16) but $[I]$
and $[O^{L,\bar L}]$. Otherwise, the perturbed theory will require the
inclusion of
additional relevant operators with corresponding coupling constants.

When $|k|$ is very large, the operator $O^{L,\bar L}$ takes the form
\begin{equation}
O^{L,\bar L}=G_{\mu\nu}~\partial x^\mu\bar\partial x^\nu,\end{equation}
where $x^\mu$ are coordinates on the group manifold $G$, whereas
\begin{equation}
G_{\mu\nu}=-\frac{k^2}{8}
L_{ab}\bar L_{\bar a\bar b}~\phi^{b\bar b}e^a_{(\mu}\bar e^{\bar
a}_{\nu)}.\end{equation}
Here $e^a_\mu$ and $\bar e^{\bar a}_\nu$ define left- and right-invariant
Killing vectors respectively. Thus, in the classical limit ($|k|\to\infty$),
the operator $O^{L,\bar L}$ becomes the nonlinear sigma model term with
metric given by eq. (25). The renormalizability of the sigma model will
allow only kinetic sigma model term and identity operator to appear on the
right hand side of the
fusion in eq. (16). In turn, the equation (23) will guarantee that this kinetic
term will
have the same structure as $O^{L,\bar L}$. Thus, in the large $|k|$ limit,
dots on the right hand side of eq. (16) can be dropped out.
Also, it is useful to point out that in the
given limit, the operator $\phi^{a\bar a}$ goes to identity. Therefore, the
operator $O^{L,\bar L}$ can be presented as a Thirring like current-current
interaction
\begin{equation}
O^{L,\bar L}\to S_{a\bar a}~J^a\bar J^{\bar a},\end{equation}
with
\begin{equation}
S_{a\bar a}=L_{ab}\bar L_{b\bar a}.\end{equation}

All in all, in the large $|k|$ limit, the operator $O^{L,\bar L}$ is suitable
for performing a renormalizable one-parametrical
perturbation around the nonunitary WZNW model. A study of such a perturbation
will be the aim of this section.

It is convenient to rescale the matrices $L,~\bar L$ as follows
\begin{equation}
L=-\frac{2}{k}\hat L,~~~~~~~~\bar L=-\frac{2}{k}\hat{\bar L}.\end{equation}
Correspondingly, eq. (23) takes the form
\begin{equation}
\hat L_{ab}=
\hat L_{ac}\hat L_{cb}~+~\frac{2}{k}\left(\hat L_{cd}\hat
L_{ef}f^{cea}f^{dfb}~+~\hat L_{cd}f^{cef}f^{dfa}
\hat L_{be}~+~\hat L_{cd}f^{cef}f^{dfb}\hat L_{ae}\right),\end{equation}
with a similar equation for $\hat{\bar L}$.

Let us consider the following theory
\begin{equation}
S(\epsilon)=S_{WZNW}(k)~-~\epsilon~\int d^2z~O^{\hat L,\hat{\bar L}}(z,\bar
z),\end{equation}
where $\epsilon$ is thought of being a small parameter. Note that $k$ is chosen
to be negative and $|k|\to\infty$. The theory is renormalizable if and only if
the matrices $\hat L,~\hat{\bar L}$ fulfill the master equation (29). Suppose
it is the case. Then to leading orders in $\epsilon$, one can write down
the renormalization group equation for the coupling
$\epsilon$. To second order, one finds (see e.g. \cite{Cardy})
\begin{equation}
{d\epsilon\over dt}\equiv\beta=
(2-2\Delta_O)\epsilon~-~\pi~C~\epsilon^2~+~{\cal O}(\epsilon^3),
\end{equation}
with $C$ being computed from the formula
\begin{equation}
\langle O^{\hat L,\hat{\bar L}}(z_1,\bar z_1)O^{\hat L,\hat{\bar L}}(z_2,\bar
z_2)O^{\hat L,\hat{\bar L}}(z_3,\bar z_3)\rangle=C||O^{\hat L,\hat{\bar
L}}||^2\Pi^3_{i<j}{1\over|z_{ij}|^{2\Delta_O}},\end{equation}
where
\begin{equation}
||O^{\hat L,\hat{\bar L}}||^2=\langle O^{\hat L,\hat{\bar L}}(1)O^{\hat
L,\hat{\bar L}}(0)\rangle.\end{equation}

By using results of \cite{Soloviev-3}, we find the coefficient $C$ to leading
order in $1/k$ is
\begin{equation}
C=\frac{1}{c_V}{\hat L_{al}\hat L_{bm}\hat L_{cn}~f^{abc}f^{lmn}~
\hat{\bar L}_{\bar a\bar l}\hat{\bar L}_{\bar b\bar m}\hat{\bar L}_{\bar c\bar
n}~f^{\bar a\bar b\bar c}f^{\bar l\bar m\bar n}\over \hat L_{dd}~\hat{\bar
L}_{\bar d\bar d}}~+~{\cal O}(1/k).\end{equation}
In this formula, the matrices $\hat L$ and $\hat{\bar L}$ are computed
perturbatively in $1/k$ \cite{Obers}. For example,
\begin{eqnarray}
\hat L_{ab}&=&\hat L^{(0)}_{ab}~+~{\cal O}(1/k),\nonumber\\ & & \\
L^{(0)}_{ab}&=&\frac{1}{2}{\sum_c}\Omega_{ac}\Omega_{bc}~\theta^c,\nonumber
\end{eqnarray}
where the quantities $\Omega_{ac}$ and $\theta^c$ are defined by
\cite{Obers}
\begin{equation}
\Omega~\Omega^{\mbox{T}}=1,~~~~~~~~
\theta^a=0~~\mbox{or}~~1,~~~~~a=1,...,\dim G.\end{equation}
Besides, there is the quantization condition \cite{Obers}
\begin{equation}
0={\sum_{cd}}\theta^c(\theta^a+\theta^b-\theta^d)~\hat f_{cda}\hat
f_{cdb},~~~~~a<b,\end{equation}
where
\begin{equation}
\hat f_{abc}=f^{lmn}~\hat L^{(0)}_{al}\hat L^{(0)}_{bm}\hat
L^{(0)}_{cn}.\end{equation}
Similarly one can find expression for $\hat{\bar L}$.

Given a solution of the master equation, we can find the constant $C$ and,
correspondingly, a fixed point $\epsilon^{*}$
of the beta function in eq. (31). This fixed
point is an infrared conformal point of the theory described by action (30).
We find
\begin{equation}
\epsilon^{*}=-{2c_V\over\pi Ck}.\end{equation}
Since different $L$'s and $\bar L$'s give rise to different $C$'s, there will
be as many infrared conformal points as $\frac{1}{2}N(N+1)$ with $N$ the total
number of solutions to the master equation.

The perturbative expression for the Virasoro central charge at the point
$\epsilon^{*}$ is given by the Cardy-Ludwig formula \cite{Cardy}
\begin{equation}
c(\epsilon^{*})=c_{WZNW}(k)-{(2-2\Delta_O)^3\over C^3}||O^{\hat L,\hat{\bar
L}}||^2,\end{equation}
where
\begin{equation}
||O^{\hat L,\hat{\bar L}}||^2={k^2\hat L_{aa}\hat{\bar L}_{\bar a\bar a}\over
4\dim G}.\end{equation}

Unfortunately, it is very difficult to identify the given perturbative
conformal points with some exact conformal theories. In particular, we could
not
clarify whether or not our action given by eq. (30) provides a Lagrangian to
the affine-Virasoro construction. But the remarkable fact that conformal points
of our theory are related to the solutions of the master Virasoro equation
makes such a hypothesis quite plausible. We believe that there is a certain
combination of the matrices $L$ and $\bar L$ which again is a solution of the
master equation. For some particular pairs of $L$ and $\bar L$ this combination
is likely to appear in the affine-Virasoro stress tensor at the corresponding
conformal point $\epsilon^{*}$.

There are two more interesting issues which we left for further
investigation. It might be interesting to use the operator $O^{L,\bar L}$ to
perturb gauged WZNW models. Another interesting question is the fusion
algebra of operators  $O^{L,\bar L}$ with different pairs of matrices
$L,~\bar L$ obeying the master equation. We hope to return to these problems
in future publications.

\par \noindent
{\em Acknowledgement}: I would like to thank S. J. Gates Jr., C. M. Hull, E.
Kiritsis and E. Verlinde for fruitful conversations. I thank also W. Sabra for
reading the manuscript.

\end{document}